\documentclass{llncs}
\usepackage{amsmath,amssymb}
\usepackage{graphicx,xspace,units,subfigure}
\usepackage[colorlinks]{hyperref}

\newcommand{\noi}{\noindent}
\newcommand{\be}{\begin{equation}}
\newcommand{\ee}{\end{equation}}

\begin{document}

\title{Reducing individuals' risk sensitiveness can promote positive and non-alarmist views about catastrophic events in an agent-based simulation}

%
%
\author{Daniele Vilone\inst{1,2} \and Francesca Giardini\inst{3} \and Mario Paolucci\inst{1}
\and Rosaria Conte\inst{1}}
%
%
%

\institute{LABSS (Laboratory of Agent Based Social Simulation),
Institute of Cognitive Science and Technology,
National Research Council (CNR), 
Via Palestro 32, 00185 Rome, Italy,\\
\and
Grupo Interdisciplinar de Sistemas Complejos (GISC), Departamento de Matem\'aticas,
Universidad Carlos III de Madrid, 28911 Legan\'es, Spain
\email{daniele.vilone@gmail.com}
\and
Faculty of Behavioral and Social Sciences, Department of Sociology, University of Groningen (Netherlands),\\
\email{f.giardini@rug.nl}}

\maketitle              

\begin{abstract}
We present a cognitive model of opinion dynamics which studies the behavior of
a population of interacting individuals in the context of risk of natural disaster.
In particular, we investigate the response of the individuals to the information
received by institutional sources about the correct behaviors for prevention and
harm reduction. The results of our study show that alarmist opinions are more likely
to be adopted by populations, since worried people tend to share their points of
view more often than other individuals.
\end{abstract}

\section{Introduction}

In the managing of natural disasters, a key passage is represented by
the implementation of prevention policies. In particular, to this aim
a fundamental task of
the official institutions is the correct communication of the proper
behaviors the population under risk should adopt before, during and
after the catastrophic event in order to minimize its negative
consequences.

In the evaluation of a risky situation, several factors, which interact
among themselves in complex ways, have to be taken into account.
In order to understand which dynamics emerge and how they act in
populations vulnerable to natural risks, it is necessary to model
these processes and test such models by means of different tools,
among them numerical simulations. 

In this paper, we are going to address the following research issues,
which are the main points for implementing the correct policies for
the cases we are dealing with:
\begin{itemize}
\item How do individuals process their opinions/beliefs about risky
events?
\item Which is the role of information from institutional sources with
respect to the information an individual collects in his/her social
groups (family, friends, colleagues?
\item Which are the collective processes of social influence by
which opinions and behaviors spread through the population?
\item How much the processes described above are in their turn
influenced by the trust the citizens yield to the institutions and
by their risk sensitiveness?
\end{itemize}

\noi In order to answer to the questions above, we have to determine the
internal mechanisms of the individuals which drive the opinion dynamics
both at microscopic and macroscopic level. In particular, we have to take
into account some important factors. Firstly, humans tend to overestimate
the risk of catastrophic but unlikely events with respect to more common
but less disastrous ones~\cite{slo82}. Secondly,
several studies verified that the trust of the individuals towards the
sources of information becomes decisive when the available information
is little~\cite{luh89,cve95}. Moreover, it was also
observed that considering situations perceived as risky (for instance, 
OGM-technologies, stocking of radioactive waste, {\it etc.}), individuals
with higher trust towards official institutions (government,
companies, scientists, {\it etc.}) repute catastrophic events less
probable~\cite{sie99,bas96}.
In short, if we want to model correctly these processes, we have to take
into account also these dynamics, together with a more realistic picture
of the mental schemes and phenomena by means of which humans elaborate
their views and beliefs~\cite{gia15}.
The model we have defined and utilized in this work is therefore
inspired from the social science point of view to the considerations above.
On the other hand, as we are going to show in the next section, it is
formally an agent-based opinion dynamics model, close to the Deffuant
model~\cite{def99,def00}, that is, a continuous-opinion kinetic exchange
model~\cite{sen14}.

The paper is organized as follows: in the next section we define the model
and how it is numerically implemented; in Section~3 we present the results
of the simulations and in Section~4 we discuss them; finally, the last
section is dedicated to the conclusions and perspectives.

\ 

\section{The Model}

We set a population of $N$ interacting agents. Each agent $i$ is characterized by an
opinion $O_i$ about the risky event, {\it i.e.}, the probability, as estimated by $i$ itself,
that the catastrophic event could actually take place.
The opinion is an external variable, that is, it can be
seen by other individuals. Besides, we define some internal variables to describe 
the internal state of agents, as well as their mental dynamics. Such internal variables
are the risk sensitivity $R_i$, the tendency
to inform others $\beta_i$, the trust towards the institution $T_i$ and the trust towards peers
$\Pi_i$. In this work, for seek of simplicity we assume that the trust towards the institution is
anti-correlated to that towards peers: $\Pi_i=1-T_i$. This is a strong but non-unrealistic
approximation~\cite{slo93}, considering that many people are suspicious of the ``official''
communications, trusting more information received by relatives, friends, or even by unknown
people on the web.

In Table~\ref{table1} the variables defining the agent internal and external behavior
are summarized, together with their main features.

\begin{table}[!ht]
\caption{
{\bf Scheme of the internal and external variables defining the agents.}}
\begin{tabular}{|c|c|c|}
\hline
\multicolumn{1}{|c|}{\bf Variable} & \multicolumn{1}{|c|}{\bf Description} & \multicolumn{1}{|c|}{\bf Notes}\\ \hline
$O_i$ & Opinion & Real number $\in[0,1]$; evolving \\ \hline
$R_i$ & Risk sensitivity & Integer $=0,\pm1$; constant \\ \hline
$\beta_i$ & Tendency to communicate & Real number $\in[0,1]$; constant \\ \hline
$T_i$ & Trust towards institution & Real number $\in[0,1]$; constant \\ \hline
$\Pi_i$ & Trust towards peers & $\Pi_i\equiv1-T_i$ \\ \hline
\end{tabular}
\label{table1}
\end{table}

In this paper we consider a mean-field approach, that is, we put our population on a complete graph: every individual can interact directly with everyone else.

\subsection{Algorithm of the dynamics}

Each time step of the simulation is made up of two stages: the institutional
communication stage, followed by a round of information exchange among peers.
More in depth, the generic $t$-th time step will take place as shown in the following.

\ 

{\it Institutional communication - } In this first stage the Institution communicates
to every player its own risk evaluation $I$. Therefore, every agent processes this
information according its opinion and internal variables. Firstly, the player $i$
modifies its old opinion $O_i(t-1)\equiv O_i^o$ following the Deffuant-like~\cite{def00} rule

\be
\label{st1a}
O_i^o\longrightarrow O_i=O_i^o+T_i(I-O_i^o) \ . 
\ee

\noi Subsequently, $O_i$ is further processed according $i$'s risk sensitivity:

\be
\label{st1b}
O_i\longrightarrow
\left\{
\begin{array}{lll}
\frac{1}{2}(1+O_i) & \ \ \ \ \  & \mbox{se} \ \ \ \ \  R_i=+1 \\
 & & \\
O_i & \ \ \ \ \   & \mbox{se} \ \ \ \ \  R_i=0 \\
 & & \\
\frac{1}{2}O_i & \ \ \ \ \   & \mbox{se} \ \ \ \ \  R_i=-1 \ . 
\end{array}
\right.
\ee

\noi More precisely, agents with positive risk sensitivity will overestimate the
institutional information, agents with negative risk sensitivity will underestimate
it, and neutral ones will not process the information further.

Once every individual has elaborated the institutional communication as described above,
the information exchange phase will take place.

\ 

{\it Information exchange among peers - } This second stage is in its turn composed by $N$
rounds. In each round a couple of agents is picked up at random. Let us call $i$ and $j$
the two agents, and $O_i$ and $O_j$ their opinions before the interaction, respectively.
Now, the probability that player $i$ ($j$) communicates its opinion to the opponent is

\be 
\label{st2a}
P_{i(j)}=O_{i(j)}^{1/\beta_{i(j)}} \ , 
\ee

\noi because we assume that given the same opinion the agents with higher tendency to
communicate are more likely to speak, but given the same tendency to communicate the
more worried agents will also speak more often.

For simplicity, let us consider $j$ as the ``speaker'' and $i$ as the ``listener''
(the symmetrical interaction where $i$ is the speaker and $j$ the listener will
take place in the same way). If the speaker decides not to give the listener its
opinion $O_j$ (according previous equation, this happens with probability $1-P_j$),
the listener's opinion $O_i$ does not change. If instead agent $j$ actually shares
its opinion, agent $i$ will change its own following another Deffuant-like rule:

\be
\label{st2b}
O_i\longrightarrow O_i'=O_i+\Pi_i(O_j-O_i)\equiv O_i'=O_i+(1-T_i)(O_j-O_i) \ . 
\ee

\noi Then, the listener processes further its new opinion again according its risk sensitivity:

\be
\label{st2c}
O_i'\longrightarrow
\left\{
\begin{array}{lll}
\frac{1}{2}(1+O_i') & \ \ \ \ \  & \mbox{se} \ \ \ \ \  R_i=+1 \\
 & & \\
O_i' & \ \ \ \ \   & \mbox{se} \ \ \ \ \  R_i=0 \\
 & & \\
\frac{1}{2}O_i' & \ \ \ \ \   & \mbox{se} \ \ \ \ \  R_i=-1 \ . 
\end{array}
\right.
\ee

\noi After $N$ rounds (so that on average each player has interacted once per time step),
the information exchange ends, and the opinions of the agents become their opinions at time
$t$.

\ 

{\it Initial conditions - } At the beginning of every simulation, the agents are randomly
assigned an opinion between 0 and 1, always with uniform distribution. Also the internal variables are randomly distributed, but the distribution is not necessarily uniform, and
will be specified in each case. We recall tha fact that whilst the opinions evolve, the
internal variable are constant in time. The institutional information $I$, which can be seen
as the ``opinion'' of the Institution, is set at the start of the dynamics and never changes.
All the simulation results are averaged over 2000 independent realizations.

\ 

{\it Asymmetrical peer interactions - } We tested also a slight modification of
the algorithm described above: in the information exchange phase, instead of picking
up $N$ times a couple of agents which share their opinions each other, we select
$2N$ times a couple made up of a fixed speaker (which only communicates its opinion
but does not listen to the opponent's one) and a fixed listener (which evolves its
opinion following the interaction, but does not share its original one). Anyway,
the two versions of the model show the very same behavior (apart very small
numerical differences), so that in the following we are reporting only the results
of the symmetrical interaction model.

\ 

\section{Results}

\subsection{Balanced systems}

Let us start our review considering a perfectly balanced system: not only all the initial opinions are uniformly distributed in the real interval $[0,1]$, but also the internal
variables $\{\beta_i, R_i, T_i\}_{i=1,\dots N}$ are picked up at random with a uniform
probability. Therefore, on average we begin with a neutral population.

\begin{figure}[!h]
  \centering
\includegraphics[width=7.1cm]{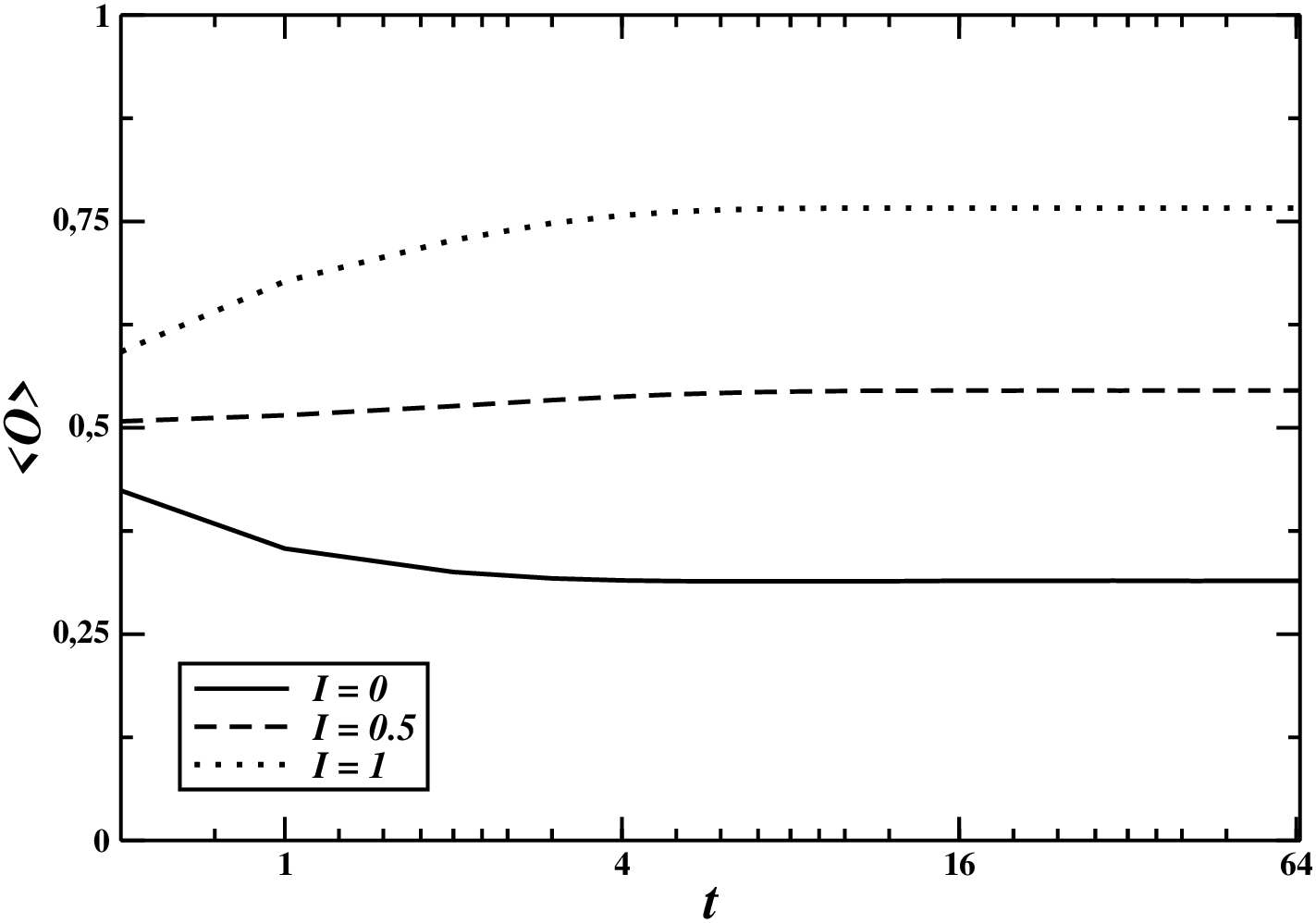}
\caption{
Time behavior of the average opinion of a system of $N=1000$ agents for totally
non-alarmist, neutral and highly alarmist institutional information. Completely
balanced population: initial opinions, trust towards institution, tendency to
speak and risk sensitivity randomly assigned with uniform distributions.}
\label{stat_dyn}
\end{figure}

First of all, we want to check if a final stationary state can be effectively reached
by the system: this is actually the case, as shown in Figure~\ref{stat_dyn}. Noticeably,
the convergence up to the final state is quite fast: already after very few time steps
the average opinion has acquired a stable value.

\begin{figure}[!h]
  \centering
\includegraphics[width=7.1cm]{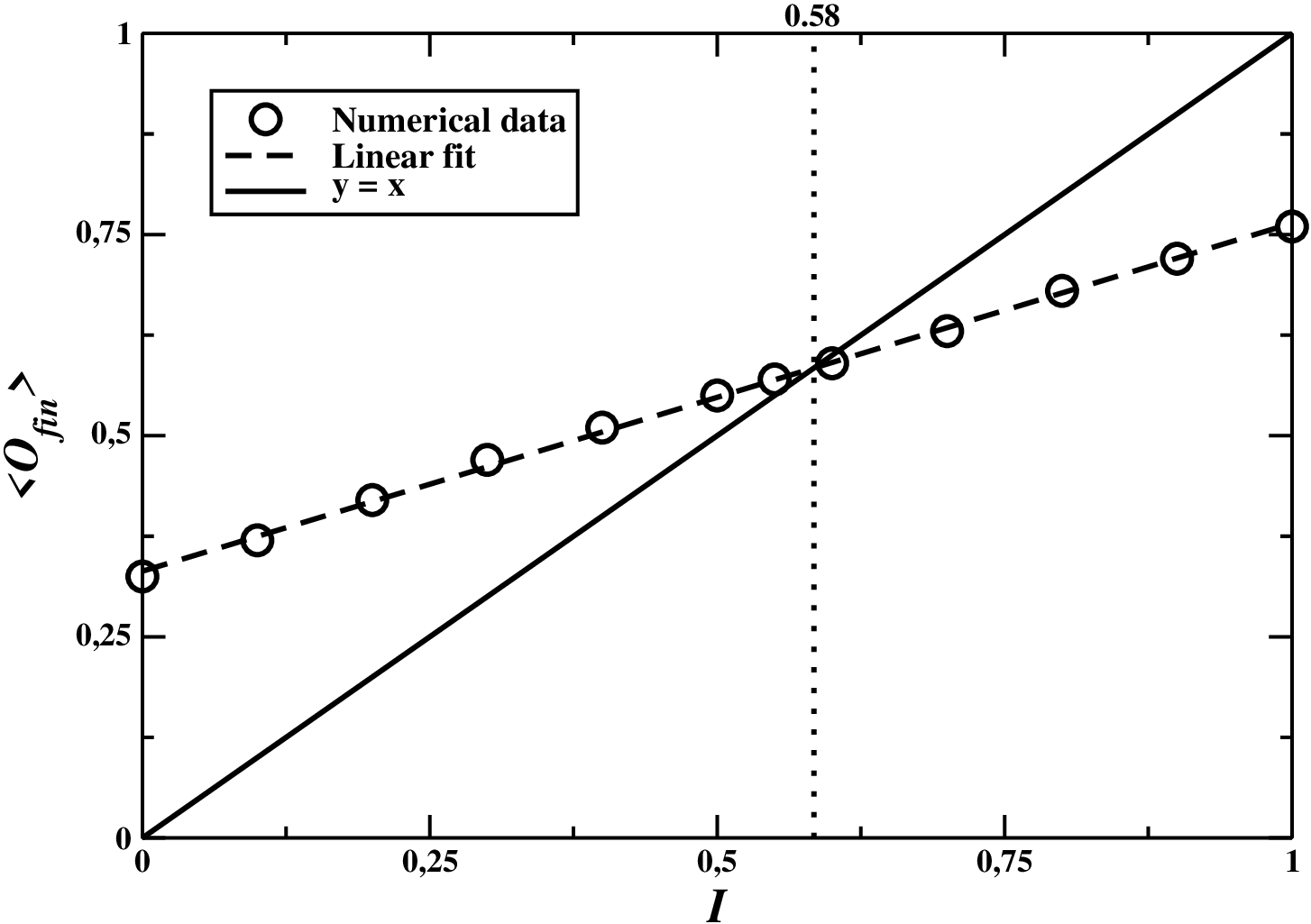}
\caption{
Behavior of the final average opinion as a function of the institutional
information for a system of $N=1000$ agents. Completely
balanced population: initial opinions, trust towards institution, tendency to
speak and risk sensitivity randomly assigned with uniform distributions. Linear
fitting parameters: intercept~$\simeq0.33$, slope~$\simeq0.43$.}
\label{stat_fin}
\end{figure}

Afterwards, it is important to understand how the populations responds to the
institutional inputs, that is, how the final average opinion behaves as a
function of the institutional information. This is shown in Figure~\ref{stat_fin}.

As it is easy to see, the system shows to be more alarmist than the institution
for appeasing information, but results less alarmed if instead the official
information is worrying. However, these results exhibit another interesting
asymmetry: the value $I^*$ of the institutional information for which the response
of the population is equal to the input is not $\bar I=0.5$, as one could expect
since the system is balanced, but larger (more precisely, we have here $I^*\simeq0.58$):
this counterintuitive outcome (in favor of alarmist opinions) is due to
the fact more alarmist people share their
opinions more often than non-alarmists, so that worried messages have more chances
to spread throughout the entire population.

\begin{figure}[!h]
\begin{centering}
\includegraphics[width=7.1cm]{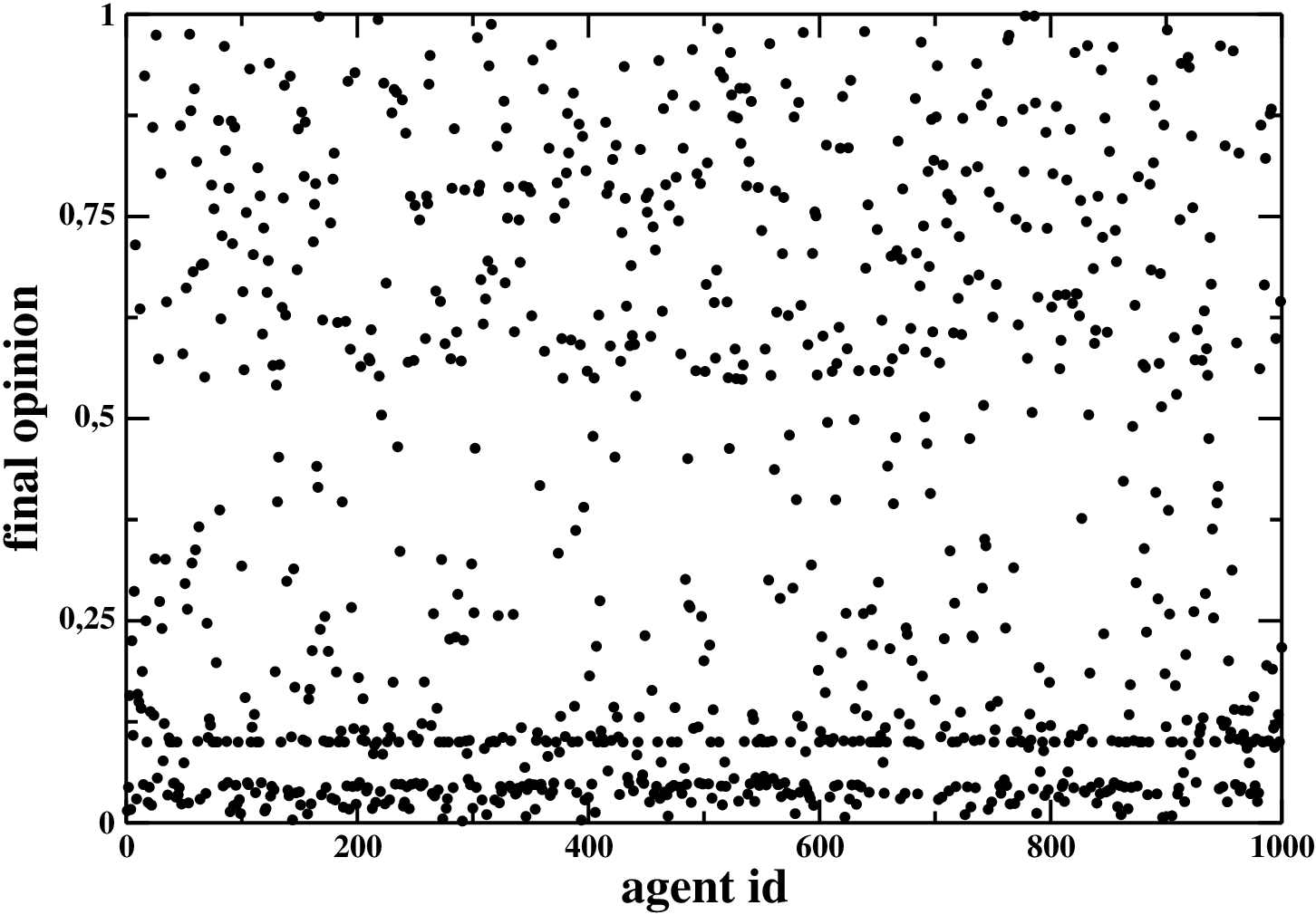}
\includegraphics[width=7.1cm]{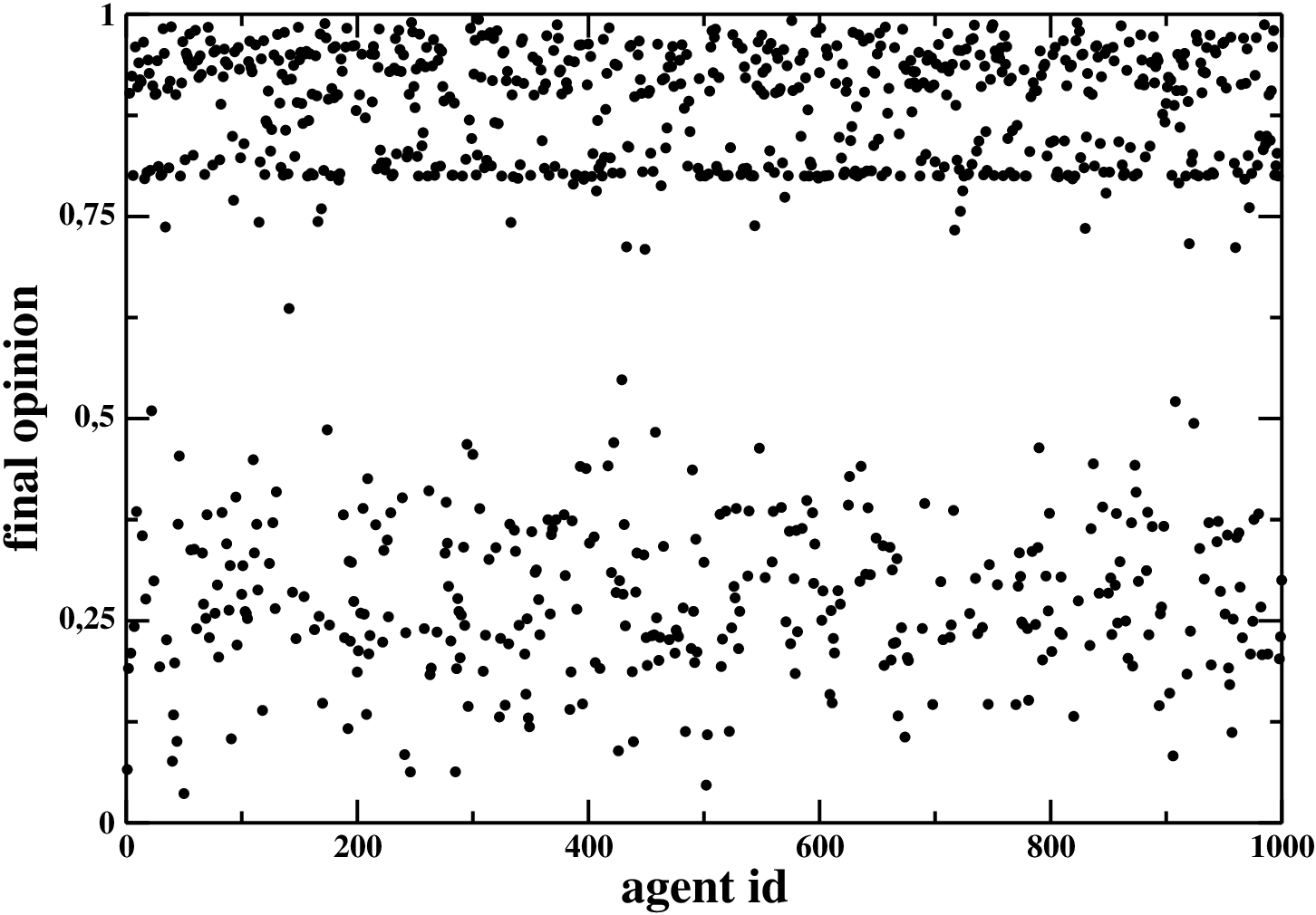}
\end{centering}
\caption{
Final opinion distribution for totally balanced systems of $N=1000$ agents
with institutional information equal to 0.10 (left) and 0.80 (right).}
\label{bal_finc}
\end{figure}

We have also to notice that in any case consensus is not reached: in the final state
it does not happen that every individual has an opinion equal to the average, but that
there is a final stationary opinion distribution, as shown in Figure~\ref{bal_finc}.
As it is easy to see, the median opinions are less common in the final configuration,
especially when the system ends up in an alarmist state. In general, even though on
average the population has a well-defined view about the risk, there are always
many contrarians which oppose the majority's opinion.

\subsection{Unbalanced systems}

In this subsection we investigate what happens when the system is not balanced,
that is, in case the distributions among agents of the internal variables are
not uniform (equivalently, their average is not equal to their median value).
In particular, we studied the behavior of the system by varying the average
risk sensitiveness and trust towards the institution.

\subsubsection{Varying trust, balanced risk sensitiveness}

Let us start with the case of systems in which the risk sensitiveness is again
uniformly distributed among agents, while the trust towards the institution
has an unbalanced distribution. More precisely, each player $i$ is assigned with
probability $P_T$ a trust $T_i$ uniformly distributed between 0.5 and 1 (high trust),
and with probability $1-P_T$ a trust $T_i$ uniformly distributed between 0 and 0.5
(low trust). Therefore, the average trust is

\be 
\langle T\rangle=\frac{1+2P_T}{4} \ .
\label{av_tr}
\ee

\noi It is worth to notice that as $P_T$ is tuned from 0 to 1, $\langle T\rangle$
goes from 0.25 to 0.75. 

\begin{figure}[!h]
  \centering
\includegraphics[width=7.1cm]{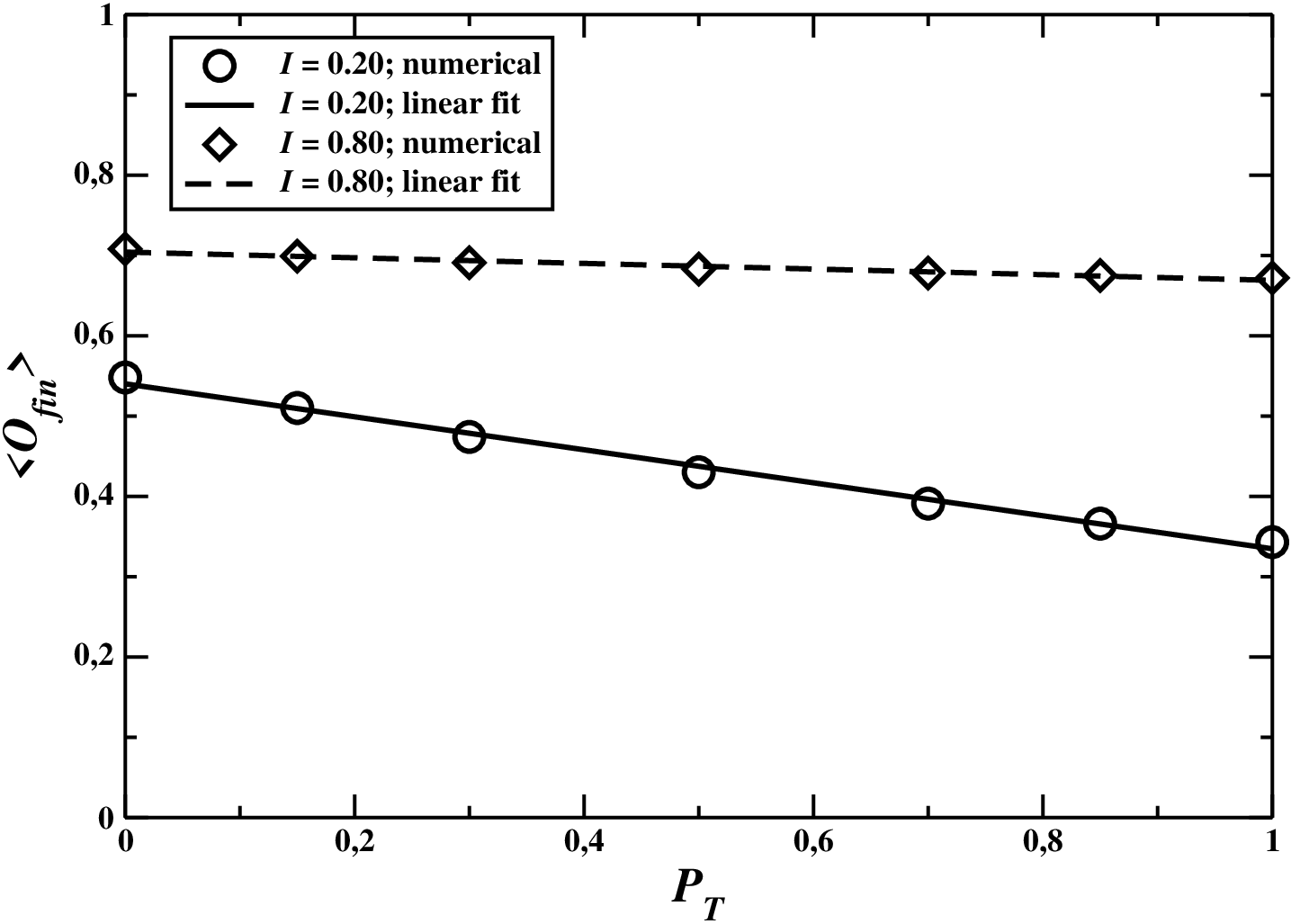}
\caption{
Behavior of the final average opinion as a function of the unbalance probability
$P_T$ of the trust towards the institution for systems of $N=1000$ agents, balanced 
risk sensitiveness, and two different values of the institutional information.
\newline Linear fitting parameters: {\bf a)} $I=0.20$, intercept~$\simeq0.54$ and
slope~$\simeq-0.21$; {\bf b)} $I=0.80$, intercept~$\simeq0.70$ and slope~$\simeq-0.03$.}
\label{unb_T}
\end{figure}

In Figure~\ref{unb_T} we show how by varying the average trust towards the
institution a population responses to the official input. Understandably,
when the institution communicates a non-alarmist message ($I=0.20$), increasing
the trust means decreasing the final average opinion. On the other hand, when
the input is alarmist ($I=0-80$) , we would expect the opposite effect: in fact,
in this case the system is much less sensitive to the trust, indeed the final
average opinion is almost constant with respect to $P_T$ (what is more, it
slightly decreases with $P_T$ increasing).

\subsubsection{Varying risk sensitiveness, balanced trust}

Here we analyze the opposite case where trust towards the institution is uniformly
distributed, but risk sensitiveness is not. In particular, every player $i$ is assigned
the neutral risk sensitiveness ($R_i=0$) with probability $1/3$, a positive risk
sensitiveness $R_i=+1$ with probability $2P_R/3$, and a negative one with
probability $2(1-P_R)/3$. Therefore, the average risk sensitiveness is

\be 
\langle R\rangle=\frac{2P_R-1}{3} \ . 
\label{av_r}
\ee

\noi In this way, as $P_T$ varies from 0 to 1, $\langle R\rangle$ goes from
$-1/3$ to $1/3$.

\begin{figure}[!h]
  \centering
\includegraphics[width=7.1cm]{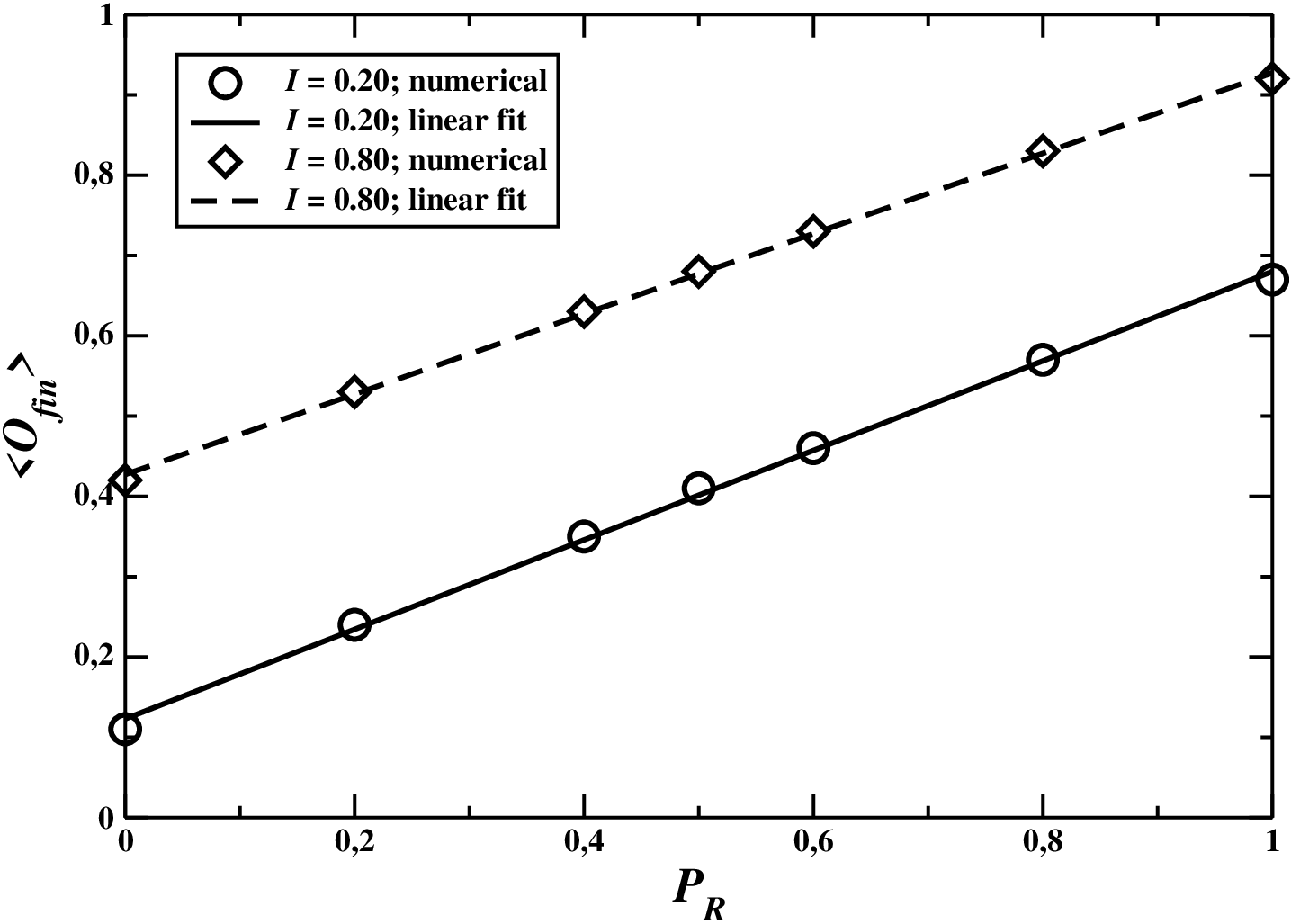}
\caption{
Behavior of the final average opinion as a function of the unbalance probability
$P_R$ of the risk sensitiveness for systems of $N=1000$ agents, balanced trust
towards the institution, and two different values of the institutional information.
\newline Linear fitting parameters: {\bf a)} $I=0.20$, intercept~$\simeq0.12$ and
slope~$\simeq0.56$; {\bf b)} $I=0.80$, intercept~$\simeq0.43$ and slope~$\simeq0.50$.}
\label{unb_R}
\end{figure}

In Figure~\ref{unb_R} we show the behavior of the final average opinion as
a function of the risk sensitivity unbalance $P_R$, again for an alarmist
institutional information ($I=0.80$) and a non-alarmist one ($I=0.20$). As
expected, $\langle O_{fin}\rangle$ increases linearly  as the population
increases its global risk sensitiveness, in the same way for different
values of $I$.

\subsection{Beyond the mean-field topology}

In the previous subsections we focused our study on systems in mean-field
approximation, that is, populations acting on complete graphs. Then, it is
also worth to understand how a change in topology affects the outcome of the
model. In order to do that, we considered the following four networks:
\begin{description}
\item[a -] a one-dimensional ring of $N=1000$ nodes with connections to
second-nearest-neighbors (so that each agent is linked to four other
individuals);
\item[b -] an Erd\"os-R\'enyi random network of $N=1000$ nodes with probability
of existence of a link $p=0.1$;
\item[c -] a Watts-Strogatz small-world network~\cite{wat98}, generated from the ring
defined above with rewiring probability $p_r=0.05$;
\item[d -] a real network of $N=1133$ users of the e-mail service of the
University of Tarragona, Spain~\cite{emnet}, which can be approximated for
high degrees with a scale-free network network with exponent $\simeq2$.
\end{description}

\begin{figure}[!h]
\begin{centering}
\includegraphics[width=7.1cm]{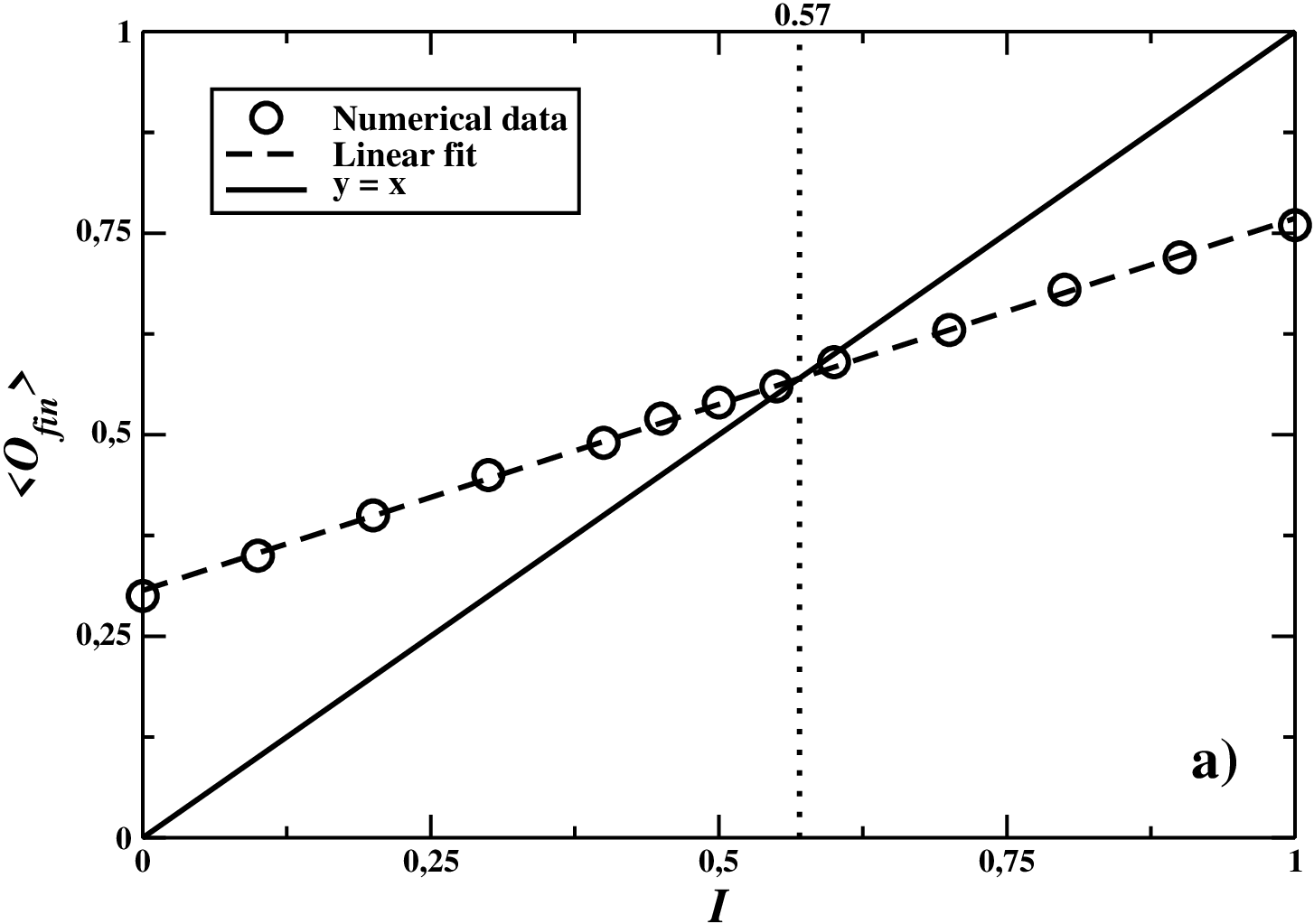}
\includegraphics[width=7.1cm]{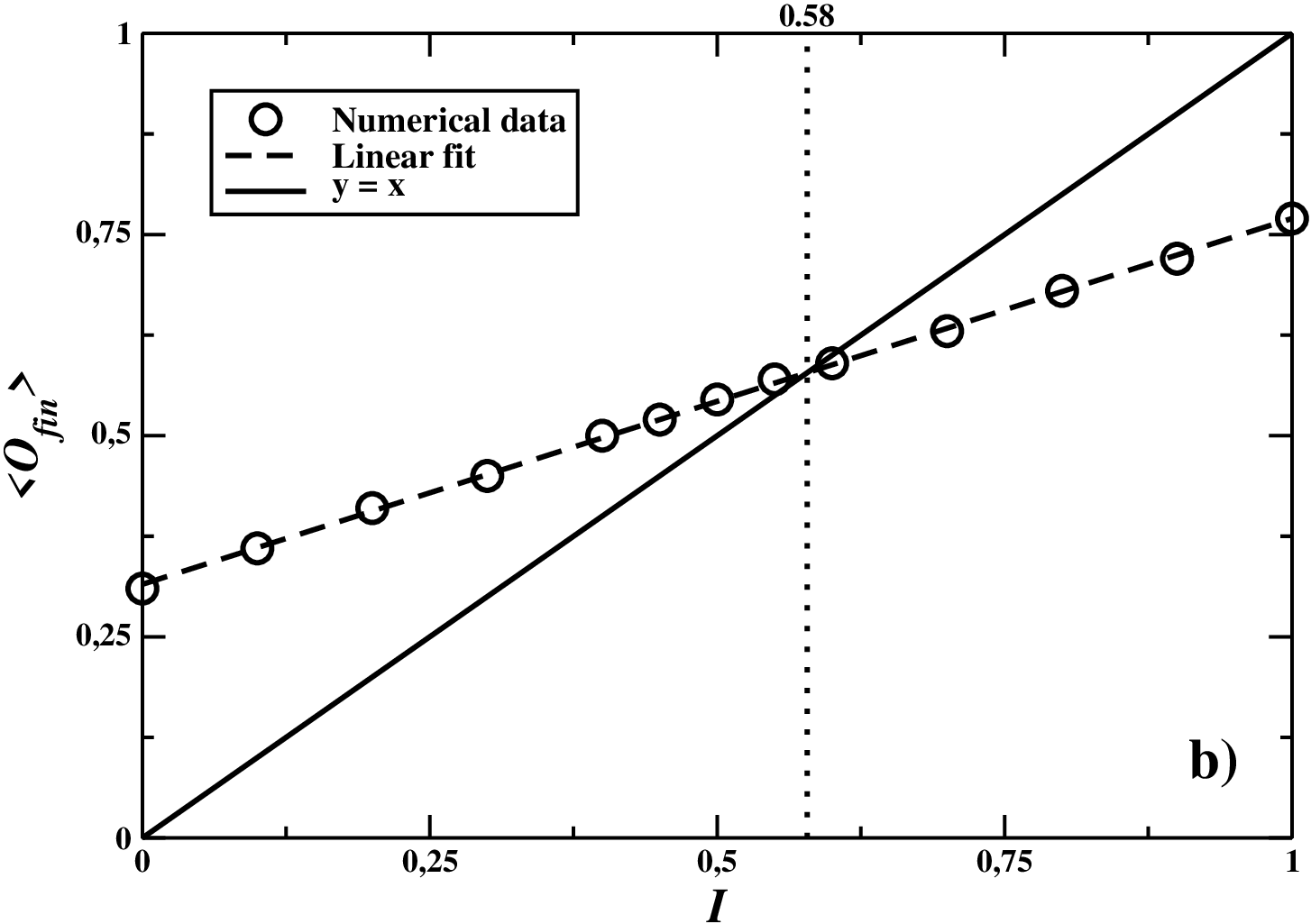}
\includegraphics[width=7.1cm]{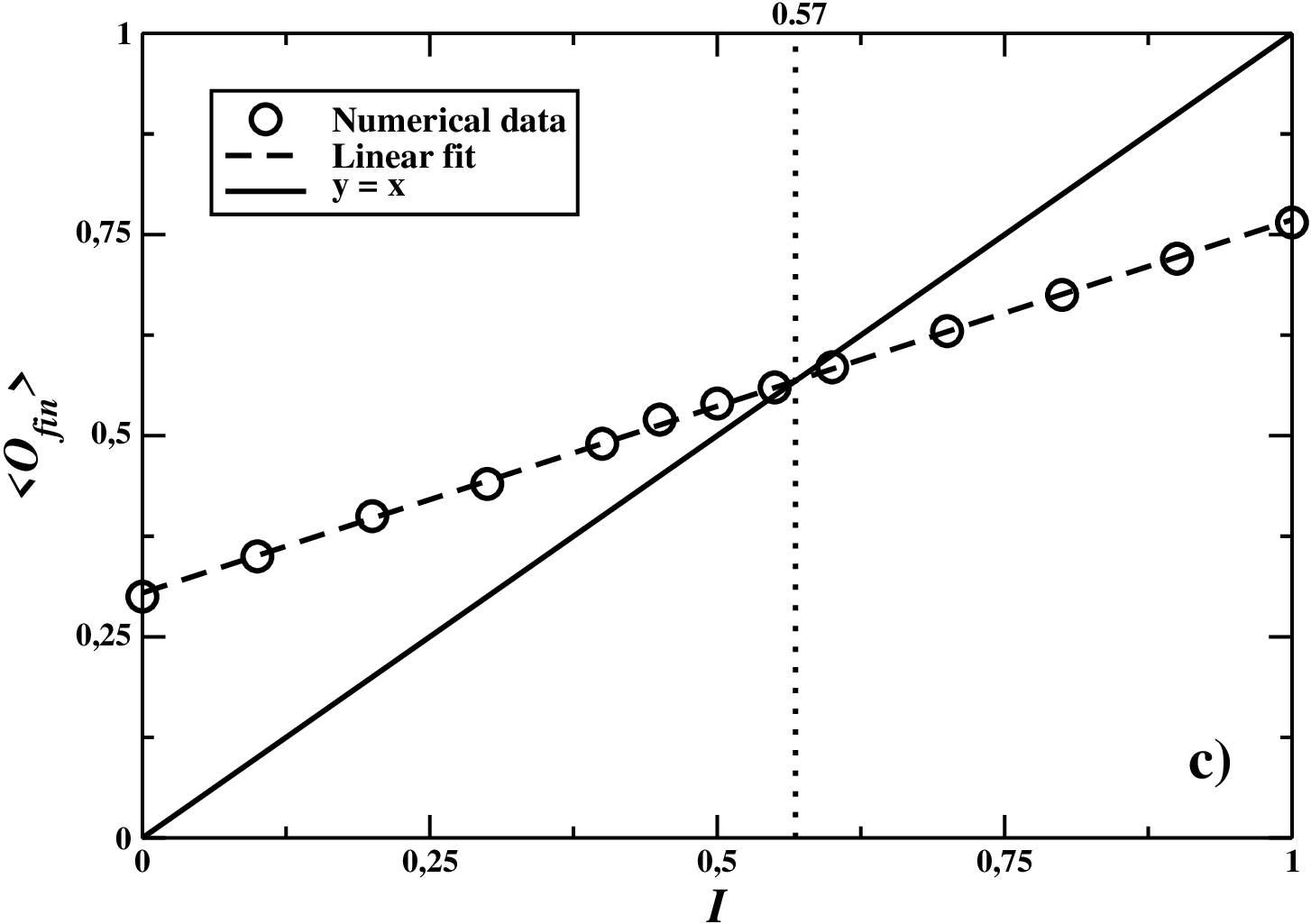}
\includegraphics[width=7.1cm]{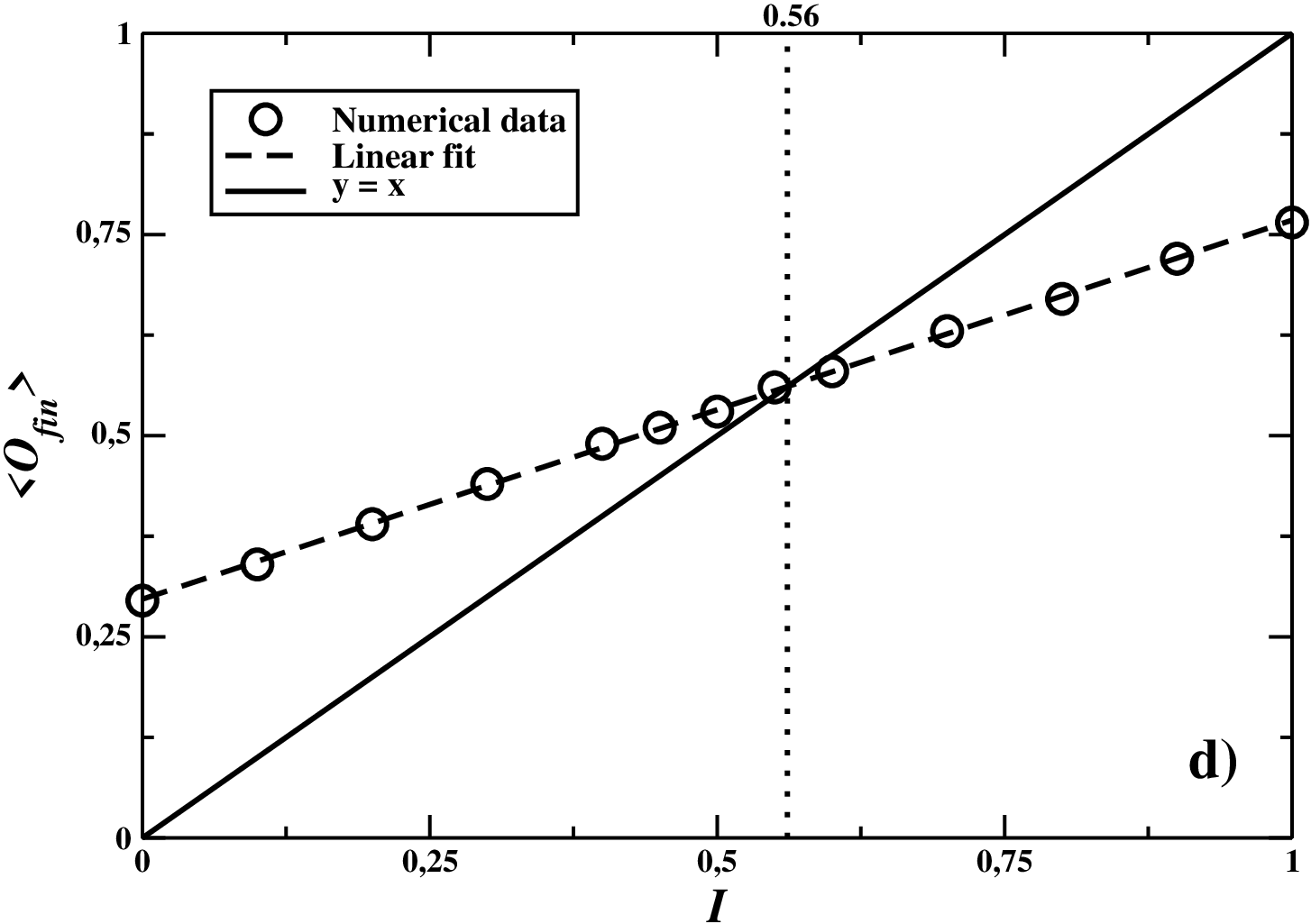}
\end{centering}
\caption{
Behavior of the final average opinion as a function of the institutional information for systems of $N=1000$ agents. Totally balanced populations.
In particular: {\bf a)} One-dimensional ring, {\bf b)} Erd\"os-R\'enyi network;
{\bf c)} Watts-Strogatz small-world network; {\bf d)} Real e-mail network.
\newline Linear fitting parameters: {\bf a)} intercept $\simeq0.31$, slope $\simeq0.46$;
{\bf b)} intercept $\simeq0.31$, slope $\simeq0.45$;
{\bf c)} intercept $\simeq0.30$, slope $\simeq0.46$;
{\bf d)} intercept $\simeq0.30$, slope $\simeq0.47$.}
\label{top_comp}
\end{figure}
As it results clear from Figure~\ref{top_comp}, the influence of the topology
is negligible, meaning that the relevant effects are due to other factors, in
particular the internal variable distributions, as we are going to discuss
in the next section.

\ 

\section{Discussion}

In this work we have simulated the dynamics of a population subject to
risk of natural disasters. Each agent gets information about the level
of the risk, that is, on the probability that a catastrophic event can
actually take place, from an institutional source and by exchanging
opinions and ideas with other agents, adopting an opinion as a consequence
of the interactions with the institution and peers. The individuals process
the information according to their mental attitudes and inclinations: in
particular, we considered three internal features: the trust towards
the institutional sources (anticorrelated to the trust towards peers),
the risk sensitiveness and the tendency to communicate. 

\ 

The main results we have obtained in our simulations are the following.
First of all, we observe that in general the global final average opinion is
more alarmist than the institutional message when the latter is reassuring,
and vice versa. This is due to the presence of agents with low trust
towards the institution, which act as ``contrarians'' in any situation.

Secondly, also in balanced populations (that is, populations where
trust, risk sensitiveness and tendency to communicate are uniformly
distributed), in the final state there is an asymmetry in favor of
the alarmist opinions: this happens because an alarmist agent will
share its opinion more often than a non-alarmist one with the same
tendency to communicate.

Concerning the role of the internal variables, it is worth to notice
that the risk sensitiveness is more influencing than the trust towards
the institution: indeed, by varying the distribution of the former
the final average opinion results much more affected than by changing
the latter. Therefore, according to these results, when the institution
transmits appeasing messages, if we want that the population  follows the
official indication, it would be better to act on the risk sensitiveness
of the citizens than on their trust.

Finally, we have checked that the topology on which the dynamics takes
place is substantially irrelevant for the final fate of the system: this
is not a complete novelty, since there are some social dynamics processes
which were experimentally shown to be independent from the details of
the networks~\cite{gru10}.

\ 

\section{Conclusions and perspectives}

In this paper, we have applied a computational approach to the study of collective risk evaluation processes. While we know, for example, that individuals in many situations tend to overestimate the risk of catastrophic events, little we know about the collective effects of those biases. We have simulated the joint effect of institutional communication with individual opinion exchange, showing how social interaction modifies the effects of institutional communication in a complex way. Individuals, exchanging opinions under a similarity bias, can polarize against institutional messages and reduce their effectiveness. Our simulations  also highlighted the prevalent role of risk sensitiveness with respect to trust, independently of connectivity. Alarmist opinions prevail in the model because they incite agents to share more. Risk sensitivity, thus, is much more effective as an intervention target with respect to risk perception. 

In the future, we plan to extend these results and to validate them. For validation ex ante, natural experiments could be searched for, that is, empirical studies where individuals are exposed to experimental and control conditions by policy choices, nature, or other factors outside the control of the investigators. On the other hand, validation ex post would benefit from a cross-methodologically approach, hybridizing simulation with experiments, online or in the laboratory. 

\ 

\section*{Acknowledgments}

Work supported by project CLARA (CLoud plAtform and smart underground imaging for natural Risk Assessment), funded by the Italian Ministry of Education and
Research (PON 2007-2013: Smart Cities and Communities and Social Innovation; Asse e Obiettivo: Asse II - Azione Integrata per la Societ\`a dell'Informazione; Ambito: Sicurezza del territorio), and by H2020 FETPROACT-GSS CIMPLEX Grant No. 641191

\ 

%
%


\begin{thebibliography}{50}
%

\bibitem{slo82}
Slovic, P., Fischoff, B., Lichtenstein, S.:
Facts versus fears (1982).

\bibitem{luh89}
Luhmann, N.:
Ecological communication. University of Chicago Press (1989).

\bibitem{cve95}
Earle, T. C., Cvetkovic, G.:
Social Trust: Towards a Cosmopolitan Society.
Westport, CT: Praeger (1995).

\bibitem{sie99}
Siegrist, M.:
A causal model explaining the perception and acceptance of gene technology 1.
J. Appl. Soc. Psych. 29(10), 195-204 (1999).

\bibitem{bas96}
Bassett, G. W., Jenkins-Smith, H. C., Silva, C.:
On-site storage of high level nuclear waste: Attitudes and perceptions of
social residents.
Risk Analysis 31(2), 309-319 (1996).

\bibitem{gia15}
Giardini, F., Vilone, D., Conte, R.:
Consensus Emerging from the
Bottom-up: the Role of Cognitive Variables in Opinion Dynamics.
Front. Phys. 3, 00064 (2015).

\bibitem{def99}
Weisbuch, G., Deffuant, G., Neau D., Amblard, F.:
Interacting agents and continuous opinions dynamics.
arXiv:cond-mat/0111494 [cond-mat.dis-nn] (2001).

\bibitem{def00}
Deffuant, G., Neau, D., Amblard, F., Weisbuch, G.:
Mixing beliefs among interacting agents.
Adv. Compl. Syst. 3, 87 (2000).

\bibitem{sen14}
P. Sen, P., Chakrabarti, B. K.:
Sociophysics - An Introduction. Oxford (2014).

\bibitem{slo93}
Slovic, P.:
Perceived risk, trust, and democracy.
Risk analysis 13(6), 675-682 (1993).

\bibitem{wat98}
Watts, D. J., Strogatz, S. H.:
Collective dynamics of 'small-world' networks.
Nature 393, 440 (1998).

\bibitem{emnet}
http://deim.urv.cat/\~{}alexandre.arenas/data/xarxes/email.zip

\bibitem{gru10}
Gruji\'c, J., Fosco, C., Araujo, L., Cuesta, J. A., S\'achez, A.:
Social experiments in the mesoscale: Humans playing a spatial prisoner’s
dilemma.
PLoS ONE 5, e13749 (2010).

\end{thebibliography}
\end{document}